\documentclass[preprint,openbib]{aastex62} % double column, single-spaced document
\usepackage{exscale}
\usepackage{amsmath}
\usepackage{color}
\usepackage{rotating}
\usepackage{graphicx}
\usepackage{epstopdf}
\usepackage{float}
\usepackage{amsfonts}
\usepackage{amssymb}
\usepackage{multirow}
\usepackage{hyperref}
\usepackage{natbib}
\usepackage{caption2}
\usepackage{subfigure}
\usepackage{overpic}
\usepackage{ulem}
\usepackage{xcolor}
\usepackage{bm}
\usepackage{booktabs}

\submitjournal{ApJS}

\shorttitle{Automatic CME Detection and Tracking with Machine Learning}
\shortauthors{Wang et al.}

\begin{document}
\correspondingauthor{Li Feng}
\email{zhangyannju@nju.edu.cn, lfeng@pmo.ac.cn}
\title{A New Automatic Tool for CME Detection and Tracking with Machine Learning Techniques}
%\author{Pengyu Wang \altaffilmark{1}, Yunyi Pan \altaffilmark{1}, Li Feng \altaffilmark{2}, Yuan Gan \altaffilmark{1},
%Yan Zhang\altaffilmark{1}, Lei Lu \altaffilmark{2} }
%\altaffiltext{1}{Department of Computer Science and Technology, Nanjing University, 210023 Nanjing, China}
%\altaffiltext{2}{Key Laboratory of Dark Matter and Space Astronomy, Purple Mountain Observatory, Chinese Academy of Sciences, 210034 Nanjing, China}

\author{Pengyu Wang}
\affil{Department of Computer Science and Technology, Nanjing University, 210023 Nanjing, China}

\author{Yan Zhang}
\affil{Department of Computer Science and Technology, Nanjing University, 210023 Nanjing, China}

\author{Li Feng}
\affil{Key Laboratory of Dark Matter and Space Astronomy, Purple Mountain Observatory, Chinese Academy of Sciences, 210034 Nanjing, China}

\author{Hanqing Yuan}
\affil{Department of Computer Science and Technology, Nanjing University, 210023 Nanjing, China}

\author{Yuan Gan}
\affil{Department of Computer Science and Technology, Nanjing University, 210023 Nanjing, China}

\author{Shuting Li}
\affil{Key Laboratory of Dark Matter and Space Astronomy, Purple Mountain Observatory, Chinese Academy of Sciences, 210034 Nanjing, China}
\affil{School of Astronomy and Space Science, University of Science and Technology of China, Hefei, Anhui 230026, China}

\author{Lei Lu}
\affil{Key Laboratory of Dark Matter and Space Astronomy, Purple Mountain Observatory, Chinese Academy of Sciences, 210034 Nanjing, China}

\author{Beili Ying}
\affil{Key Laboratory of Dark Matter and Space Astronomy, Purple Mountain Observatory, Chinese Academy of Sciences, 210034 Nanjing, China}
\affil{School of Astronomy and Space Science, University of Science and Technology of China, Hefei, Anhui 230026, China}

\author{Weiqun Gan}
\affil{Key Laboratory of Dark Matter and Space Astronomy, Purple Mountain Observatory, Chinese Academy of Sciences, 210034 Nanjing, China}

\author{Hui Li}
\affil{Key Laboratory of Dark Matter and Space Astronomy, Purple Mountain Observatory, Chinese Academy of Sciences, 210034 Nanjing, China}

\begin{abstract}
With the accumulation of big data of CME observations by coronagraphs, automatic detection and tracking of CMEs has proven to be crucial. The excellent performance of convolutional neural network in image classification, object detection and other computer vision tasks motivates us to apply it to CME detection and tracking as well. We have developed a new tool for CME Automatic detection and tracking with MachinE Learning (CAMEL) techniques. The system is a three-module pipeline. It is first a supervised image classification problem. We solve it by training a neural network LeNet with training labels obtained from an existing CME catalog. Those images containing CME structures are flagged as CME images. Next, to identify the CME region in each CME-flagged image, we use deep descriptor transforming to localize the common object in an image set. A following step is to apply the graph cut technique to finely tune the detected CME region.  To track the CME in an image sequence,  the binary images with detected CME pixels are converted from cartesian to polar coordinate. A CME event is labeled if it can move in at least two frames and reach the edge of coronagraph field of view. For each event, a few fundamental parameters are derived. The results of four representative CMEs with various characteristics are presented and compared with those from four existing automatic and manual catalogs. We find that CAMEL can detect more complete and weaker structures, and has better performance to catch a CME as early as possible. 
  
\end{abstract}


\section{Introduction}
Observations of coronal mass ejections (CMEs) by space missions can be dated back to 1970s. The coronagraphs aboard 
\textit{Solar and Heliospheric Observatory} (SOHO) have made tremendous contributions to CME observations. Large Angle and Spectrometric Coronagraph Experiment (LASCO, \citeauthor{Brueckner:etal:1995} \citeyear{Brueckner:etal:1995}) can follow CMEs from 1.1 to about 30 $\mathrm{R_S}$.  Since the launch of the \textit{Solar TErrestrial RElations Observatory} (STEREO) mission, CMEs can be observed from two different perspectives with the coronagraphs COR 1 and COR 2 in the Sun Earth Connection Coronal and Heliospheric Investigation (SECCHI, \citeauthor{Howard:etal:2008} \citeyear{Howard:etal:2008}) instrument package. With the accumulation of big data of coronagraph images, it becomes more and more important to have the capability of automatic detection and tracking of different features, especially CMEs, and build corresponding event catalogs. On one hand, they provide much easier access to data for statistical studies of CME key parameters. On the other hand, with automatic detection the coronagraph images with CME flagged can be used immediately for the purpose of near-real-time space weather predictions.

Different CME catalogs have been developed with the long-running coronagraph observations. They are classified as either manual or automated catalogs. The manual catalog that we mostly use is the CME catalog \footnote{\url{http://cdaw.gsfc.nasa.gov/CME_list/index.html}} created for LASCO observations and maintained at the Coordinated Data Analysis Workshops (CDAW) data center \citep{yashiro2004catalog}. Event movies of observations by LASCO and other related instruments together with key parameters of each CME are provided. Although the CDAW catalog has been widely adopted, the CME detection and tracking are done by human perception and are obviously subjective and time consuming. Depending on the experience of different operators, we may reach different detection results and physical parameters. When the Sun approaches its activity maximum, the detection and tracking of CMEs require a lot of man power. 

These disadvantages of manual CME catalogs prompt the development of automatic catalogs. Several methods have been devised and deployed for the LASCO and/or SECCHI coronagraph images, for instance, Solar Eruptive Event Detection System (SEEDS, \citeauthor{Olmedo2008Automatic} \citeyear{Olmedo2008Automatic})\footnote{\url{http://spaceweather.gmu.edu/seeds/}}, Computer-Aided CME Tracking (CACTus, \citeauthor{Robbrecht:Berghmans:2004} \citeyear{Robbrecht:Berghmans:2004}; \citeauthor{robbrecht2009automated} \citeyear{robbrecht2009automated})\footnote{\url{http://sidc.oma.be/cactus/}}, CORonal IMage Process (CORIMP, \citeauthor{byrne2012automatic} \citeyear{byrne2012automatic})\footnote{\url{http://alshamess.ifa.hawaii.edu/CORIMP/}}, Automatic Recognition of Transient Events and Marseilles Inventory from Synoptic maps (ARTEMIS, \citeauthor{Boursier:etal:2009} \citeyear{Boursier:etal:2009})\footnote{\url{http://cesam.lam.fr/lascomission/ARTEMIS/}}. Thanks to the observations of two \textit{STEREO} spacecraft, dual-viewpoint CME catalogs have also been developed, e.g., \citet{Vourlidas:etal:2017}\footnote{\url{http://solar.jhuapl.edu/Data-Products/COR-CME-Catalog.php}}, for \textit{STEREO}/COR2 observations. For all the aforementioned catalogs, CMEs are detected automatically using different traditional segmentation methods. 

Nowadays machine learning technique becomes more and more widely used in many different research fields. It brings a cross-disciplinary research community together between computer science and solar and heliospheric physics. There have been quite a few applications of machine learning technique for different solar features and space weather purposes. For example, \citet{Dhuri:etal:2019} used machine learning to understand the underlying mechanisms governing flares.  \citet{Huang:etal:2018} applied deep learning method to flare forecasting.  \citet{Camporeale:etal:2017} and \citet{Delouille:etal:2018} used the machine learning technique for the classification of solar wind and coronal hole, respectively. Very recently, \citet{Galvez:etal:2019} even compiled a curated dataset for the \textit{Solar Dynamics Observatory} (SDO) mission in a format suitable for the  booming machine learning research. A review paper on the challenge of machine learning in space weather nowcasting and forecasting can be found in \citet{Camporeale:etal:2019}. 

In the field of computer vision,  machine learning has shown excellent performance in image classification, feature detection and tracking \citep{krizhevsky2012imagenet, Shelhamer2017FCN, he2017mask}. In view of its great success and our need of fast detection and tracking for CME prediction, we have developed and validated our machine learning technique CAMEL for the automatic CME detection and tracking based on the LASCO C2 data. Section 2 describes the detailed mathematical methodology including image classification, CME detection, and CME tracking. In Section 3, we compare our results with those derived from existing SEEDS, CACTus, CORIMP catalogs for four representative CMEs with different angular width, velocity, and brightness. The method is developed and tested using the observations around the solar maximum during which the CME occurrence rate is much higher than that around the solar minimum. The large number of CMEs around the solar maximum poses a challenge in CME detection and tracking. The last section is dedicated to conclusions and discussions.

\section{Methodology}
Our goal is to detect and track pixel-level CME regions in a set of white-light coronagraph images by using machine learning methods. To this end, we design a three-module algorithm pipeline. In the first module, we use a trained convolutional neural network to classify whether a coronagraph image contains CME structures or not. The images with CME structures are flagged as CME images. On the contrary, the rest images are flagged as non-CME images. The second module is to detect pixel-level CME regions in CME-flagged images by using an unsupervised common object co-localization method, and the detected CME regions are future refined using the graph-cut method in computer vision. The final module serves to track a CME in running-difference images. 

\subsection{Preprocessing}
Before we going through the pipeline, all coronagraph data are processed in the following way: the downloaded level 0.5 LASCO C2 fits files are read with $lasco\_readfits.pro$ from the Solar Software (SSW), and then processed to level 1 data using $reduce\_level\_1.pro$ from SSW. The processing consists the calibrations for dark current, flat field, stray light, distortion, vignetting, photometry, time and position correction. After the processing, the solar north has been rotated to the image north. For CME detection and tracking, we have used the running difference images as inputs to the three-module algorithm pipeline. 

As a preprocessing step, all input LASCO C2 images with 1024 $\times$ 1024 resolution are first down-sampled to 512 $\times$ 512 resolution and aligned according to coordinates of solar centers. Then all down-sampled images are passed through a noise filter to suppress some sharp noise features. In our method, we use normalized box filter with a sliding window of size 3$\times$ 3. Normalized box filtering is a basic linear image filtering, which computes the average value of surrounding pixels. Then the running-difference images are computed simply by using the following equation: 
\begin{equation}
\bm{u_}{i}=\bm{n_}{i}-\bm{n_}{i-1},
\end{equation}
where $\bm{u_}{i}$ is the running-difference image, which equals the current image $\bm{n_}{i}$ minus the previous image $\bm{n_}{i-1}$.  
For some of the LASCO images containing missing blocks, we create a missing-block mask according to the previous image: if the value of a pixel in the previous image is zero, then the same pixel of the running-difference image has also a zero value. The final running-difference image is multiplied by the missing-block mask. 

For the first module of our algorithm pipeline, we need to train a convolutional neural network (CNN) for image classification and rough localization. From the perspective of computational efficiency, our CNN takes 112 $\times$ 112 resolution images as input. After rough localization, the down-sampled images CME region will be refined by the graph cut method in the original 512 $\times$ 512 running-difference images.

\subsection{Image classification}
Detecting and locating instances of a certain class in an image can be seen as a basic problem in computer vision. Each certain class has its own special features, which can be manually or automatically extracted by a supervised machine learning method. Recently, CNNs have shown excellent performance in image classification, object detection and some other computer vision tasks. The multi-scale and sliding window approach can help CNNs learn more robust features of a certain class without any human effort and prior knowledge. Before detecting the CME events in each image, we need a CNN model to tell us if there are CME structures in every input LASCO-C2 running-difference image first. To train such a convolutional neural network in a supervised fashion, we need to first collect images and training labels. As a preprocessing step, all input running-difference with 1024 $\times$ 1024 resolution are down-sampled to 112 $\times$ 112 resolution. The training set of data are ten-month LASCO C2 images from January to October 2011 around the solar maximum whose category label is known. Both image categories flagged with or without CMEs are obtained from the CDAW catalog. The first step of CME detection can be treated as a supervised image classification problem by assigning a given white-light coronagraph image to CME-detected category or CME-not-detected category. As a second step, the middle-level features extracted from the well-trained CNN can be used for detecting the CME regions in Section 2.3. 

\begin{figure}[htbp]
   \centering
   \includegraphics[trim=.1cm .1cm 0cm 0.1cm, clip, width=0.8\textwidth]{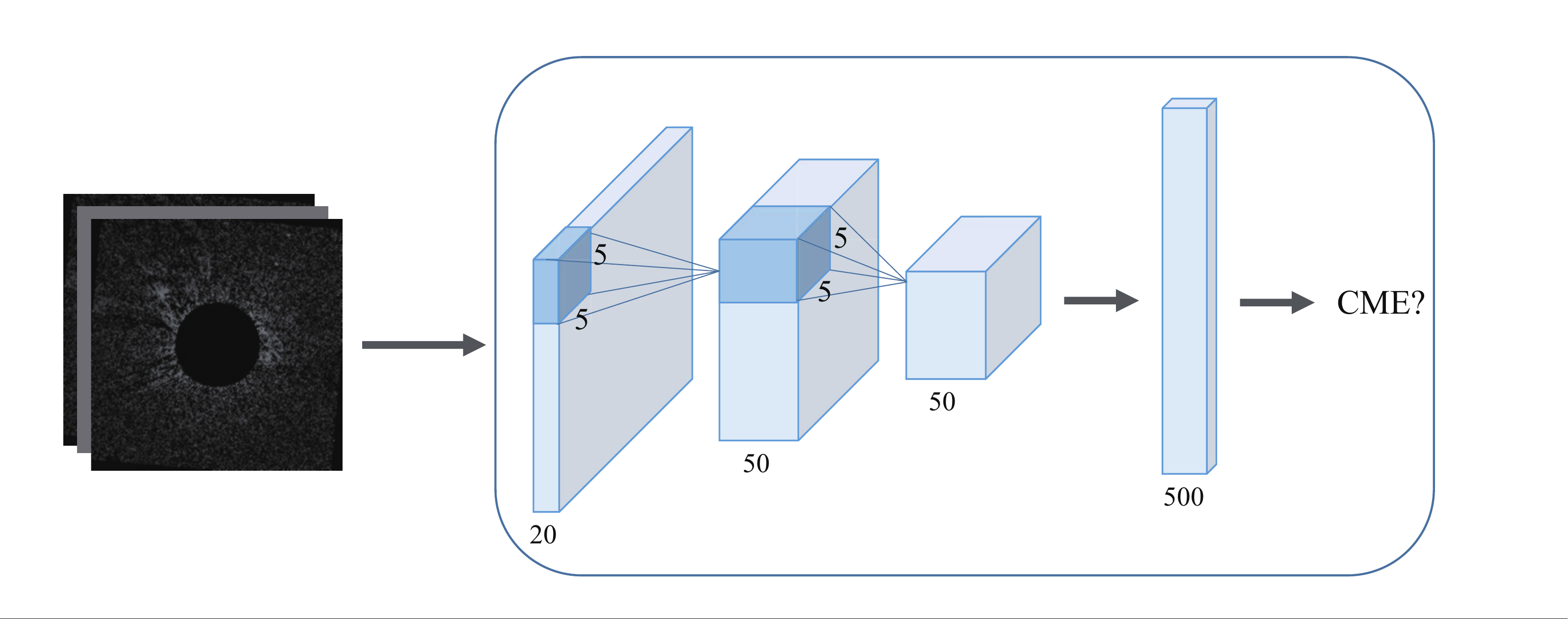}
   \caption{LeNet architecture which contains two convolution layers, two down-sampling layers and two fully-connected Layers is used for image classification. The first convolution layer has 20 5$\times$5 kernels and the second convolution layer has 50 5$\times$5 kernels. After convolution layers and down-sampling layers, the feature map of each image is down-sampled to 25 $\times$ 25 resolution. The middle-level feature map then pass through two fully connected layers and output the final CME occurrence probability.}
   \label{fig:lenet}
\end{figure}

The CNN architecture we used is called LeNet-5 from \cite{Lecun1998LeNetCNN}, which has two convolution layers,  two down-sampling layers and two fully-connected Layers. This classical architecture can be divided into two modules: a feature extractor module and a classifier module. The feature extractor module consists of convolution layers, nonlinearity activation layers and down-sampling layers and the last two fully connected layers form the classification module.  A convolution layer can be seen as a local-connected network in which each hidden unit will connect to only a small contiguous region of the input and obtain different feature activation values at each location. The convolution kernel slides from left to right and from top to bottom on the input feature map of the upper layer. Each time it slides to a position, the convolution kernel is multiplied and summed with the pixel value of input feature map block at that position, and the summation result is passed through the activation function to obtain an output pixel value of the feature map of the layer. The $j$th feature map of layer $l$ is obtained as follows:
\begin{equation}
\bm{x_}{j}^{l}=f(\displaystyle\sum_{i}^{N}\bm{x_}{i}^{l-1}*\bm{k_}{i}^{l}+\bm{b_}{j}^{l}),
\end{equation}
\begin{equation}
f(x)=max(0,x),
\end{equation}
$N$ denotes the number of feature maps of layer $l-1$, $\bm{k}$ represents convolution kernels and $\bm{b}$ is a bias term. $f$ represents the nonlinearity activation function. We use Rectified Linear Units (ReLUs), which can make the CNN training several times faster. Down-sampling layers could help to enlarge the receptive field and aggregate features at various locations. This layer treats each feature map separately. It computes the max value over a neighborhood in each feature map. The neighborhoods are stepped by a stride whose size is two.

After convolution layers and down-sampling layers, the feature map of each image is down-sampled to 25 $\times$ 25 resolution. Then the high-level semantic knowledge can be obtained via fully connected layers and output the final CME occurrence probability. The original LeNet architecture is designed for handwritten digit number recognition and the output layer outputs 10 units which represent the probability of each class (0-9). We modified the output layer and output 2 unit which represent the probability of the CME occurrence. To obtain the probability, we use two-way softmax function to  produce a distribution over the two-class labels:
\begin{equation}
P_\mathrm{{CME}}=\frac{e^{x_\mathrm{CME}}}{e^{x_\mathrm{CME}}+e^{x_\mathrm{non-CME}}},
\end{equation}
where $\mathrm{x_{CME}}$ and $\mathrm{x_{non-CME}}$ are both output units from the final output layer.  An image with the output probability value greater than 0.5 can be seen as a CME-detected image. Figure~\ref{fig:lenet} shows the LeNet architecture we use. 

As a machine learning approach, the CNN model needs to discover weights and biases automatically with training data and labels. As a classification problem, the objective loss function can be define as follows:
\begin{equation}
L=-\frac{1}{N}\displaystyle\sum_{i}^{N}y_{i}^{*}*ln(y_{i}),
\end{equation}
where $N$ denotes the number of training data, $y_{i}^{*}$ is the true label which equals 0 or 1, $y_{i}$ is the CNN output probability which is less than 1 and greater than 0. We see that $L$ is non-negative, so the aim of CNN training process is to minimize $L$ as a function of the weights and biases. We trained our models using stochastic gradient descent with a batch size of 128 examples. The update rules of weights and biases can be seen in the following:
\begin{equation}
\bm{k_}{i+1} := \bm{k_}{i} - \eta * \frac{\partial L}{\partial \bm{k_}{i}},
\end{equation}
\begin{equation}
\bm{b_}{i+1} := \bm{b_}{i} - \eta * \frac{\partial L}{\partial \bm{b_}{i}},
\end{equation}
where $i$ is the iteration index, $\eta$ is the learning rate. The learning rate was initialized at 0.0001 and reduced three times prior to termination. Only a batch of training examples are used for updating the weights and biases in each iteration. The weights in each layer are initialized from a zero-mean Gaussian distribution with a standard deviation of 0.01, and the neuron biases are initialized with a constant value of zero in each convolutional layer and fully-connected layer. In the test phase, continuous running-difference images are classified in a chronological order. A set of continuous CME-detected frames can be seen as an image sequence of CME evolution which is used for CME co-localization and tracking.

\subsection{CME detection}
\subsubsection{CME region co-localization}
After the classification, the next step is to segment the CME regions in every CME-detected image. However, due to the lack of a set of images with known labeled CME regions, we need to solve the problem in an unsupervised fashion. After training the above LeNet neural network, we can extract convolutional feature maps from the last convolution layer of the CNN model. Each feature map is considered as a down-sample of input and contains high level semantic information. To mine the hidden information for segmenting the CME regions, we use Deep Descriptor Transforming (DDT, \citeauthor{Wei2019DDT} \citeyear{Wei2019DDT}), an unsupervised image co-localization method, which utilize the Principal Component Analysis (PCA, \citeauthor{Karlpearson1901LIII} \citeyear{Karlpearson1901LIII}) to analyze CNN feature maps and localize category-consistent regions of each image in an image set. The extracted feature maps can be considered as having 25 $\times$ 25 cells and each cell contains one d-dimensional feature vector. PCA uses an orthogonal transformation to convert d-dimensional correlated variables into a set of linearly uncorrelated variables which called principal components by the eigendecomposition of the co-variance matrix . The co-variance matrix of the input data is calculated by:
\begin{equation}
Cov(x)=\frac{1}{K}\displaystyle\sum_n^{N}\displaystyle\sum_{(i,j)}^{h,w}(\bm{x_}{(i,j)}^n-\bm{\bar{x}})(\bm{x_}{(i,j)}^n-\bm{\bar{x}})^T,
\end{equation}
\begin{equation}
\bm{\bar{x}}=\frac{1}{K}\displaystyle\sum_n^{N}\displaystyle\sum_{(i,j)}^{h,w}\bm{x_}{(i,j)}^n,
\end{equation}
where $K = h\times{w}\times{N}$. $N$ denotes the number of input feature maps with $h\times w$ resolution and $\bm{x_}{(i,j)}^n$ represents a $C$ dimension CNN feature vector of image n at pixel position $(i,j)$. After the eigendecomposition, we can get the eigenvectors $\bm{\xi_}{(1)}$,$\ldots$,$\bm{\xi_}{(d)}$ of the co-variance matrix which correspond to the sorted eigenvalues $\lambda_{1}\geq \cdots\geq\lambda_{d}\geq0$ in a descending order. We take the first eigenvector which corresponds to the largest eigenvalue as the main projection direction. For a particular position $(i,j)$ of a CNN feature vector of image n, its main principal feature is calculated as follows:
\begin{equation}
f_{(i,j)}^n=\bm{\xi_}{(1)}^T(\bm{x_}{(i,j)}^n-\bm{\bar{x}}).
\end{equation}

In this way, we reduce its feature dimension from $C$ to $1$ and the feature value after transformation can be treated as the appearance possibility of the common object at each pixel position. All pixel locations of $f_{(i,j)}$ form into an indicator matrix whose dimensions are h$\times$w: 
\begin{equation}
F=
\begin{bmatrix}
 f_{(1,1)} & f_{(1,2)} & \ldots & f_{(1,w)} \\ 
 f_{(2,1)} & f_{(2,2)} & \ldots & f_{(2,w)} \\ 
 \vdots & \vdots & \ddots & \vdots \\ 
 f_{(h,1)} & f_{(h,2)} & \ldots & f_{(h,w)} \\ 
\end{bmatrix}
\end{equation}

\begin{figure}[htbp]
	\centering
	\includegraphics[width=0.8\textwidth]{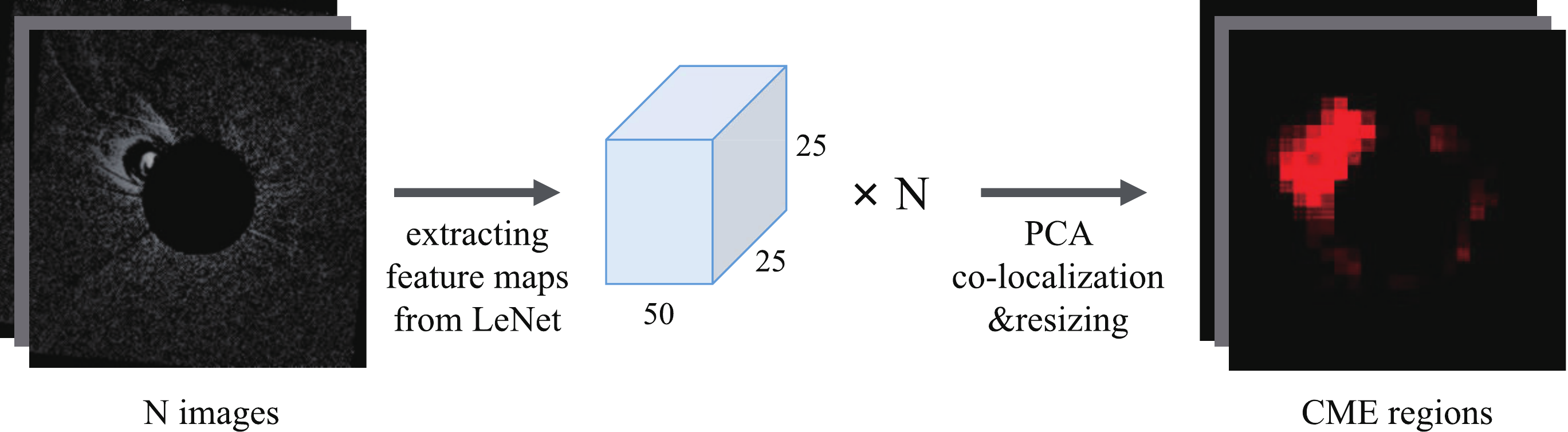}
	\caption{The pipeline of image co-localization. We extract CNN feature maps of each CME-detected image from our trained LeNet model and use PCA as projection directions for transforming these features to evaluate their correlations. }
	\label{fig:colocalization}
\end{figure}

The pipeline of image co-localization can be found in Figure~\ref{fig:colocalization}. The image sequence of CME evolution we obtain from the trained CNN model consists a set of CME images, which is directly processed through DDT algorithm for CME region co-localization. The final output of the image co-localization is a set of CME region mask images with the same resolution as that of the input feature maps. For convenience, we resize the output in the same resolution by the nearest interpolation. 

\subsubsection{CME region refinement}
\begin{figure}[htbp]
	\centering
	\includegraphics[width=0.8\textwidth]{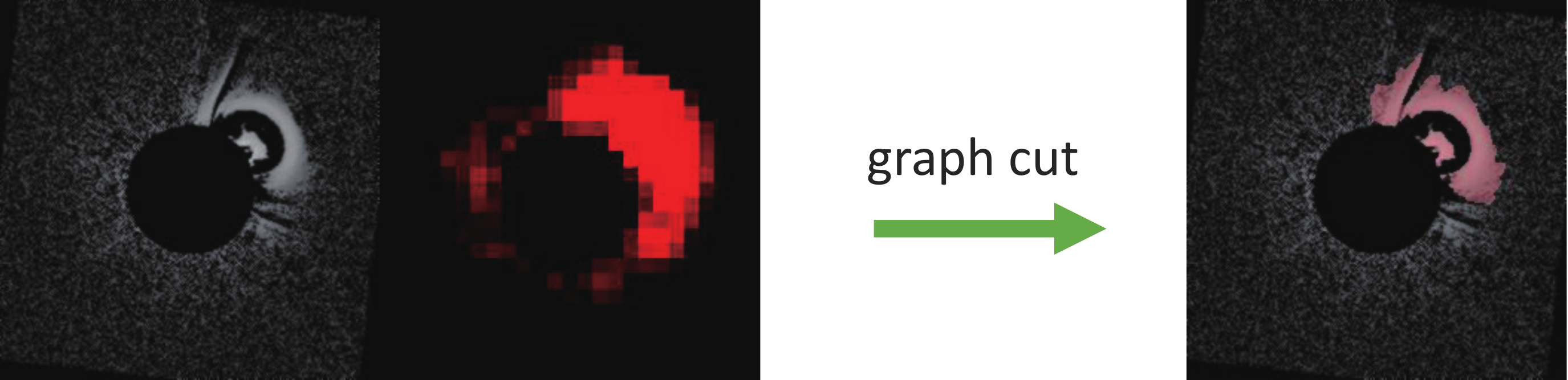}
	\caption{One example of the CME region refinement result. To get high-resolution images with CME region well annotated, we use graph cut method in computer vision which finely tunes the CME boundary.}
	\label{fig:graphcut}
\end{figure}

The outputs of the pipeline in Figure~\ref{fig:colocalization} are just images with roughly detected CME regions. To obtain images with CME region finely tuned, we use the graph cut method \citep{Boykov2001GC} in computer vision for segmented region smoothing. Obviously, the indicator matrix can only roughly tell the probabilities that a pixel position belongs to CME or non-CME class. However, class consistency may arise among neighboring pixels. To address this problem, a framework of energy minimization is naturally formulated. In this framework, one seeks the labeling $l$ of image pixels that minimizes the energy:
\begin{equation}
E(l)=\lambda_{s}E_{smooth}(l)+\lambda_{d}E_{data}(l),
\end{equation}
%\end{equation}
where $\lambda_{s}, \lambda_{d}$ are nonnegative constants to balance the influences of each term. $E_{smooth}(l)$ measures the class consistency of $l$ among boundary pixels according to their neighborhood intensity difference. While $E_{data}(l)$ measures the disagreement between $l$ and the predicted data which is optimized mainly based on the probability calculated in Section 2.2.  We set $E_{smooth}(l)$ and $E_{data}(l)$ as follows:
\begin{equation}
E_{smooth}(l)=\displaystyle\sum_{(p,q)\in{N_{8}}, lp\ne lq}e^{-\alpha(I_{p}-I_{q})^2},
\end{equation}
\begin{equation}
E_{data}(l)=\displaystyle\sum_p-log(pr(l_p)),
\end{equation}
where $pr(l_p)$ denotes the probability of a pixel position $p$ assigned as class CME and $I_{p}$ denotes the intensity on position i. The graph cut optimization can be then employed to efficiently solve the energy minimization problem.  The graph cut algorithm generates related graph of the labeling problem and the minimum cut set is used to solve the minimum cut set of that graph. Then the minimum solution of the energy function is obtained. More details of graph cut algorithm can be found in \citet{Boykov2001GC}. Here we show one example of the comparison results before and after optimization in Figure~\ref{fig:graphcut}. 

\subsection{CME tracking}
After CME region detection and refinement, we can only obtain CME regions in each frame independently. Furthermore, there could be more than one CME in the image sequence of CME evolution which we obtained in Sections 2.2 and 2.3. To track a CME in a series of running-difference images, we define some rules to identify a CME which are similar to \citet{Olmedo2008Automatic}. First, a CME must be seen to move outward in at least two running-difference images. Second, the maximal height of a CME must reach out of the C2 FOV. Each tracked CME which does not satisfy the above two rules are abandoned. Moreover, given a set of images with pixel-level CME region annotated, we aim to recognize each CME in that image set and analyze its key parameters, e.g., central position angle, angular width,  median and maximal velocities.

\begin{figure}[htbp]
	\subfigure[]{
	\begin{minipage}[t]{\linewidth}
	\centering
	\includegraphics[width=0.7\textwidth]{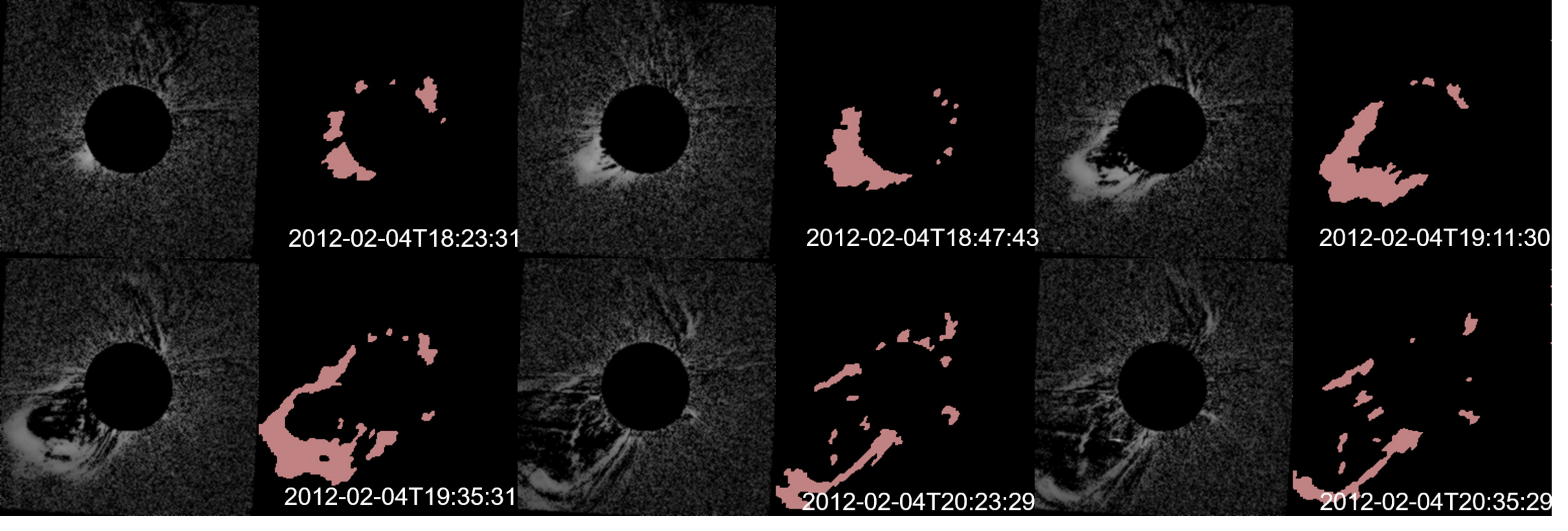}
	\end{minipage}
	\label{fig:tracking_a}
	}
	
	\subfigure[]{
	\begin{minipage}[t]{0.5\linewidth}
	\centering
	\includegraphics[width=0.9\textwidth]{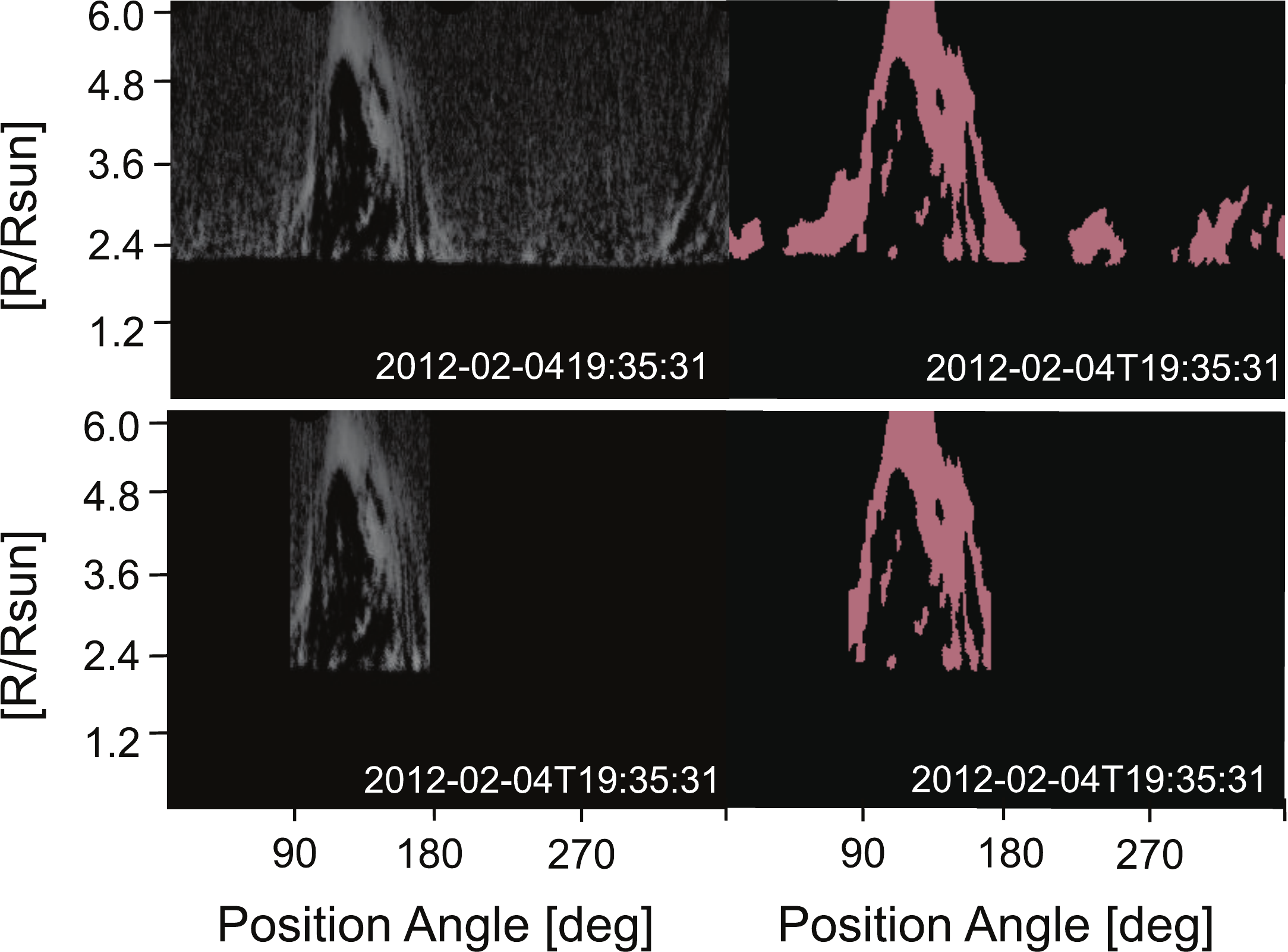}
	\end{minipage}
	\label{fig:tracking_b}
	}
	\subfigure[]{
	\begin{minipage}[t]{0.5\linewidth}
	\centering
	\includegraphics[width=1.0\textwidth]{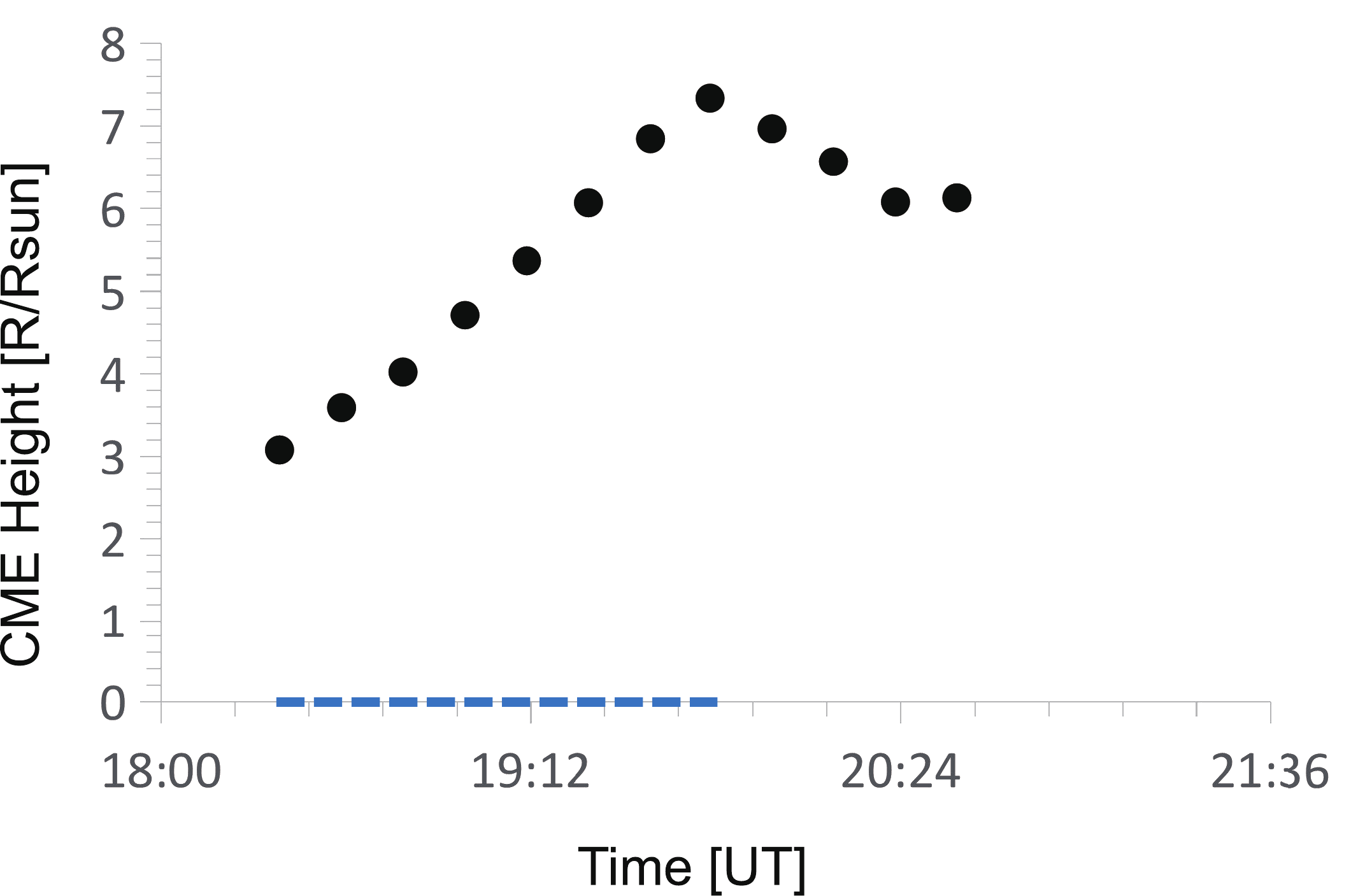}
	\end{minipage}
	\label{fig:tracking_c}
	}
	\caption{A demonstration of the CME tracking module. (a) Input of the tracking module: an image sequence of CME evolution which consists a set of CME-class images in gray color and with the pixel-level CME region annotated in red color. The selected time range is on February 4 2012. (b) The upper panel is the CME image at 19:35 UT transformed to the polar coordinate system and converted to a binary file. The lower panel is the cleaned image for the calculation of CME parameters. (c) CME height-time diagram at the position angle with the maximal velocity, in which the blue dashed line indicates the obtained the time range for the tracked CME.}
\end{figure}

For better tracking the moving of a CME, all images with CME region annotated are transformed to a polar coordinate system with $360 \times 360$ resolution. The height range at each angle position is from 2.2 to 6.2 solar radii.  As a demonstration, we use the CME event occurred on February 4, 2012. Figure~\ref{fig:tracking_a} shows the input of our tracking module which is an image sequence of CME evolution for a given time range. All images are ordered according to the observation time. The original CME images are shown in gray and the images with detected CME pixels are indicated in red.  The upper panel of Figure~\ref{fig:tracking_b} presents the results of coordinate transform for the frame at 19:35~UT as an example. Actually we apply the coordinate transformation to each image in the sequence, and compute the maximal height of the CME region mask at each position angle in the given time range. All position angles with a maximal height less than half of the FOV are removed. The rest position angles are merged together according to the position connectivity. Again using the frame at 19:35~UT as an illustration,  we show the cleaned result in the lower panel of Figure~\ref{fig:tracking_b} from where we determine the start position angle (start PA) and the end position angle (end PA) of each CME. The central position angle (PA) is the average of the start PA and the end PA. And the angular width is derived as the difference of these two PAs.

The height-time diagram at each position angle between the derived start and end position angles can be retrieved. Subsequently, the corresponding velocity can be obtained.  In Figure~\ref{fig:tracking_c}, as a representative case we plot the CME height evolution at the position angle with the maximal velocity. To determine the start and end time of each CME, we find all time segments with increasing CME height in the height-time diagram. Next, we check if the CME in each time segment meets the defined two criteria: existing in at least two frames and reaching beyond the LASCO C2 FOV. The time segment which does not satisfy the aforementioned criteria is discarded. For the case in Figure~\ref{fig:tracking_c}, the derived final time range of the tracked CME is indicated by the blue dashed line. To derive representative CME velocities, we compute the median and maximal values of the CME velocity distribution at all derived position angle. The velocity at each position angle is calculated by a linear fit to the data points in the obtained time segment. As in the CACTus catalog \citep{robbrecht2009automated}, we use the median velocity as an overall velocity of the detected CME. Meanwhile, in order to compare with the velocity in the CDAW catalog, we also calculate the maximal velocity. In summary, for a tracked CME we offer the following five fundamental parameters: first appearance time in the LASCO C2 FOV (T$_{\mathrm{start}}$), central position angle (CPA), angular width (AW), median and maximal velocities (V$_{\mathrm{med}}$ and V$_{\mathrm{max}}$). These fundamental parameters are used for the comparison among different detection and tracking techniques.

\section{Results and Comparisons} 

\begin{table*}[htbp]
\centering
 \caption{Fundamental parameters of four representative CME events observed by LASCO-C2. T$_ {\mathrm{start}}$ (UT): time of first appearance in the C2 FOV; CPA (degree): central position angle which is measured counterclockwise from the north of the Sun;  AW (degree): angular width; and CME velocity (km~s$^{-1}$), where SEEDS shows the half-max-lead velocity, CDAW shows the maximal velocity (V$_{\mathrm{med}}$) and other methods show the median and maximal velocities (V$_{\mathrm{med}}$ and V$_{\mathrm{max}}$).}
 \label{table1}
 \begin{tabular}{cccccc}
\toprule 
method &parameters& 01 Jan. 2012& 22 Nov. 2011&18 Jan. 2012&21 Mar. 2012\\
\midrule  
  \multirow{3}{*}{CAMEL}&T$_ {\mathrm{start}}$ &01:25&21:17&12:22&07:35\\
 &CPA&41&313&163&halo\\
 &AW&16&100&182&360\\
 &V$_{\mathrm{med}}$&584&440&256&633\\
 &V$_{\mathrm{max}}$&680&524&327&1042\\
  \hline 
  \multirow{3}{*}{CACTus}&T$_ {\mathrm{start}}$ &01:25&21:28&12:24&07:48\\
 &CPA&42&291&180&336\\
 &AW&18&168&98&118\\
 &V$_{\mathrm{med}}$&719&512&299&868\\
 &V$_{\mathrm{max}}$&893&702&363&1838\\
 \hline 
  \multirow{3}{*}{CORIMP}&T$_ {\mathrm{start}}$ &-&-&14:00&08:00\\
 &CPA&-&-&162&255\\
 &AW&-&-&12&3\\
 &V$_{\mathrm{med}}$&-&-&241&555\\
 &V$_{\mathrm{max}}$&-&-&-&921\\
 \hline 
  \multirow{3}{*}{SEEDS}&T$_ {\mathrm{start}}$ &01:36&21:17&14:00&08:00\\
 &CPA&61&324&177&321\\
 &AW&53&98&79&74\\
 &Vel&189&325&296&812\\
 \hline 
  \multirow{3}{*}{CDAW}&T$_ {\mathrm{start}}$ &01:25&20:57&12:24&07:36\\
 &CPA&40&291&172&halo\\
 &AW&23&157&203&360\\
 &V$_{\mathrm{max}}$&801&668&267&1178\\
\bottomrule
\end{tabular}
\end{table*}

By applying the method described above, we have detected and tracked the CMEs from the LASCO C2 running-difference images  in the time range from November 2011 to April 2012. To evaluate the performance of our machine learning techniques, we compare our results to those with some other existing automatic detection and tracking techniques, namely, CACTus, CORIMP, and SEEDs. 
Table ~\ref{table1} presents the comparison results of a few fundamental CME parameters: T$_ {\mathrm{start}}$,  CPA, AW, and velocity between our CAMEL technique, CACTus, CORIMP, SEEDS, and CDAW. Note that in Table~\ref{table1} we include the available median and maximal velocities derived with CAMEL, CACTus, and CORIMP, and include only maximal velocity derived with CDAW. The velocity derived with SEEDs is calculated from the CME height at half-max-lead \citep{Olmedo2008Automatic}.  

According to the morphological characteristics of CMEs resulted from the CAMEL detection, we have chosen four representative CME events with different angular widths. They are a jet-like narrow CME on 01 January 2012 with an AW of 16 degrees, a limb CME on 22 November 2011 with an AW of 100 degrees, a partial halo CME on 18 January 2012 with an AW of 182 degrees, and a full halo CME on 21 March 2012. On the other hand, the selected four CMEs span a large range of CME velocity from about 300 km s$^{-1}$ to more than 1000 km s$^{-1}$, and cover structures with different level of brightness.  Because CAMEL detects and tracks features which moves outwards, at the moment it does not differentiate whether a structure belongs to a CME or a CME-driven shock automatically. Therefore for the full halo CME on 22 November 2011 we actually detect and track the CME-driven shock. It has also been claimed by \citet{Kwon:etal:2014} that the halo CME they studied is primarily the projection of the bubble-shaped shock wave but not the underlying CME flux rope. Nevertheless, it nicely proves that our CAMEL method is able to detect and track not only bright but also weak signals, e.g., in this case a fainter shock wave or in other cases weak CMEs. 
From the comparison of the fundamental CME parameters, we can see that the results with our CAMEL method for these four CME events are in general more similar to the CDAW manual measurements. Systematically we find that the velocity from CACTus is always the largest, and the velocity from SEEDs is always the smallest. This is because the CME detection with CACTus is based on the J-map which has the capability to track weak signals at the CME leading edge, whereas the velocity derived with SEEDs is calculated from the CME height at the half-max-lead. As the CME velocity generally increases from inside to leading front \citep{Feng:etal:2015, Ying:etal:2019}, it is expected that CACTus yields relatively higher velocities and SEEDS produces relatively lower velocities.

\begin{figure}[htbp]
	\centering
	\includegraphics[width=0.98\textwidth]{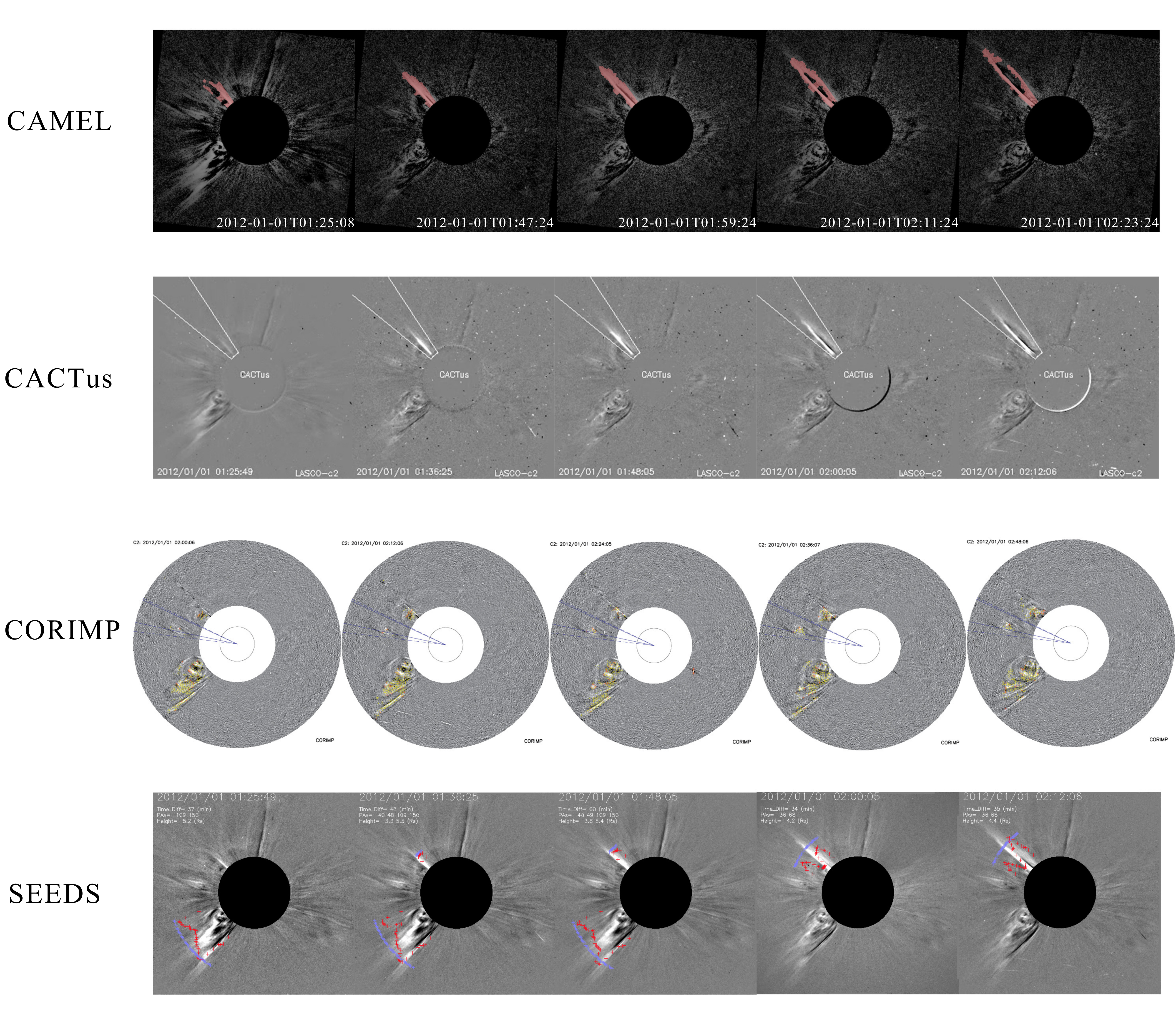}
	\caption{Comparison of the detection and tracking results for the CME event on 1 January 2012. From top to bottom rows are some selected frames showing the detection results with CAMEL, CACTus, CORIMP, and SEEDS, respectively.}
	\label{fig:compare_1}
\end{figure}

\begin{figure}[htbp]
	\centering
	\includegraphics[width=0.98\textwidth]{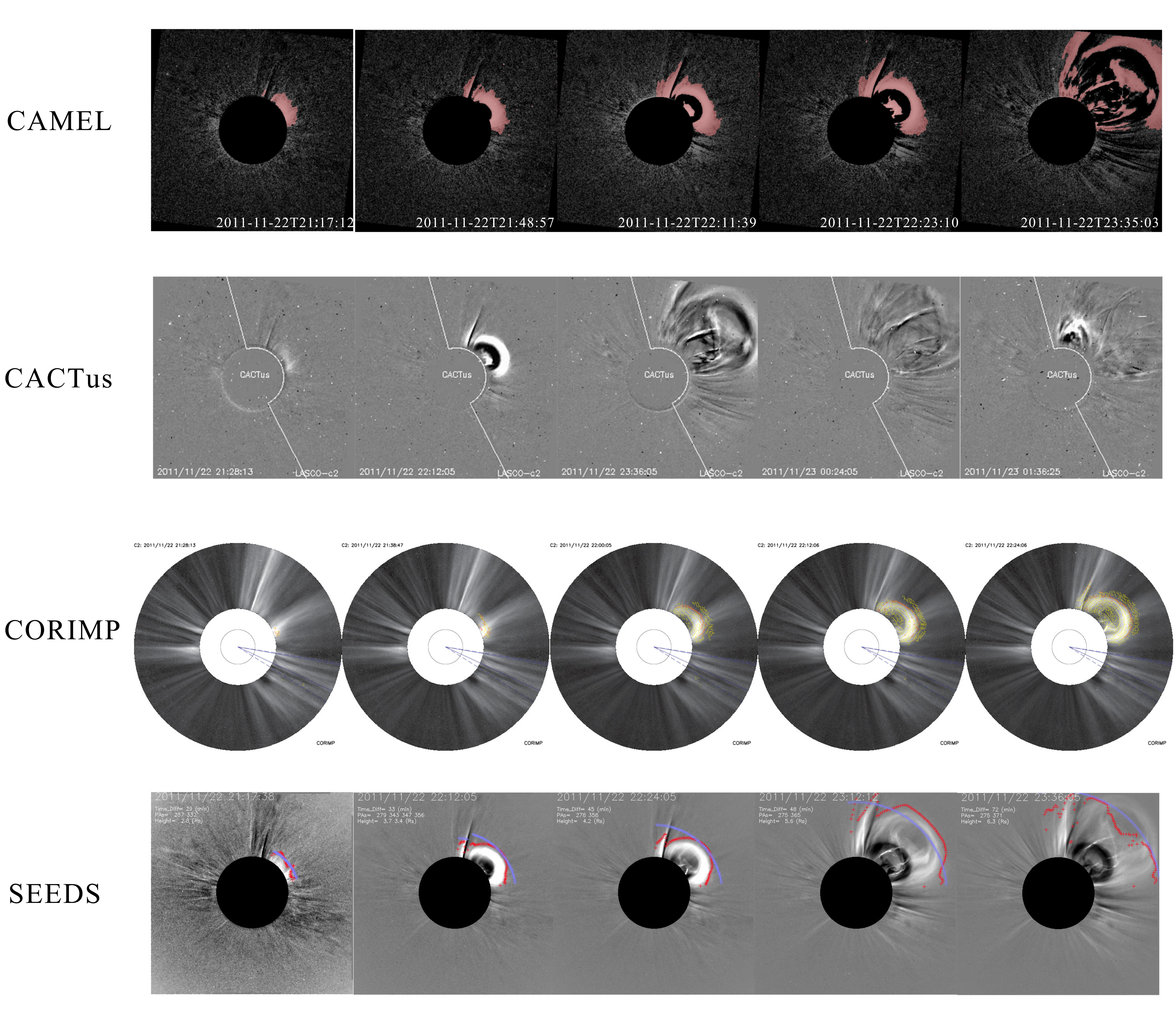}
	\caption{Comparison of the detection and tracking results for the CME event on 22 November 2011. }
	\label{fig:compare_2}
\end{figure}

In the following subsections we present individually the observations of the selected four CME events and show the comparison of the detected CME region with CAMEL, CACTus, CORIMP, and SEEDS. Note that CME parameters in Table ~\ref{table1} and the corresponding frames in Figure~\ref{fig:compare_1} to \ref{fig:compare_4} derived with CACTus, CORIMP, and SEEDS are adopted from their websites introduced in Section 1. Because the LASCO C2 data we use are processed to level 1 with the observation time corrected,  there might be up to two-minute time difference for a given frame between our results and the others. For the C2 observations on 01 January 2012, and 18 January 2012, actually besides the CME we are interested in we detect more than one CMEs in the same frame.
In such cases, the other CMEs are not annotated in pink color and each CME is grouped separately into an image sequence.

\subsection{Jet-like narrow CME on 1 January 2012}

The CME under investigation is jet-like and has a very narrow AW in the north-east quadrant of the C2 image. The detection and tracking results with CAMEL, CACTus, CORIMP, and SEEDS are displayed from top to bottom in Figure~\ref{fig:compare_1}. 
CAMEL detects its first appearance in C2 at $01:25$~UT. It is consistent with the time given by CACTus and CDAW. SEEDS detects the first appearance about 11 minutes later in the next C2 frame.  This CME is not included in the CORIMP catalog, although CORIMP recognizes it with the identified leading edge much lower than the true position. Apparently it registers another CME as marked by the blue lines in the third row of Figure~\ref{fig:compare_1}. Concerning the CPA and AW, CAMEL, CACTus, and CDAW yield consistent results with the CPA about 40 degrees and the AW about 20 degrees. SEEDS has a somewhat higher CPA of 61 degrees and wider AW of 53 degrees. For the velocity, as described before CACTus and SEEDS have the highest and lowest values respectively, CAMEL and CDAW sit in between. When we inspect the detected CME region obtained with CAMEL in pink, CORIMP in yellow and SEEDS in red,  we find that the detection with CAMEL is closest to our visual perception. CACTus does not provide the detected pixel-level positions of the CME front in its catalog. As shown in the second row, only the angular range of the CME is indicated by the white solid lines.

\subsection{Limb CME on 22 November 2011}
Limb CMEs probably belong to the most common category of CMEs. Usually different detection and tracking methods work better for this category. Figure~\ref{fig:compare_2} presents some selected frames of the detection and tracking results with CAMEL, CACTus, CORIMP, and SEEDS from top to bottom. As demonstrated in Figure~\ref{fig:compare_2}, all four automatic tools produce reasonable results. The issue with CORIMP is that although the CME we are interested in is correctly identified in the C2 FOV and its boundary appears to be well defined, it is not registered in the catalog. On the contrary, the other much weaker CME is included as indicated by the blue lines in the third row of Figure~\ref{fig:compare_2}. The first appearance time of the CME obtained by all the automatic methods is later than the time inspected by eye from the manual CDAW catalog. The CPA has a relatively small range from 291 to 324 degrees with an average of 305 degrees, while the AW has a much larger range from 98 to 168 degrees with an average of 131 degrees. The reason for this wide AW range may be partially due to the influence of the deflected streamers adjacent to the CME. It seems to be a common issue for all automatic tools to precisely separate streamers from the CME itself. As can be seen in Figure~\ref{fig:compare_2}, the deflected streamers in the north of the CME are all falsely included as part of the CME. For CACTus, the deflected streamers in the south are also included and make its AW the largest. Concerning the velocities, as expected CACTus has the highest value, SEEDS has the lowest value, CAMEL and CDAW sit in between.

\begin{figure}[htbp]
	\centering
	\includegraphics[width=0.98\textwidth]{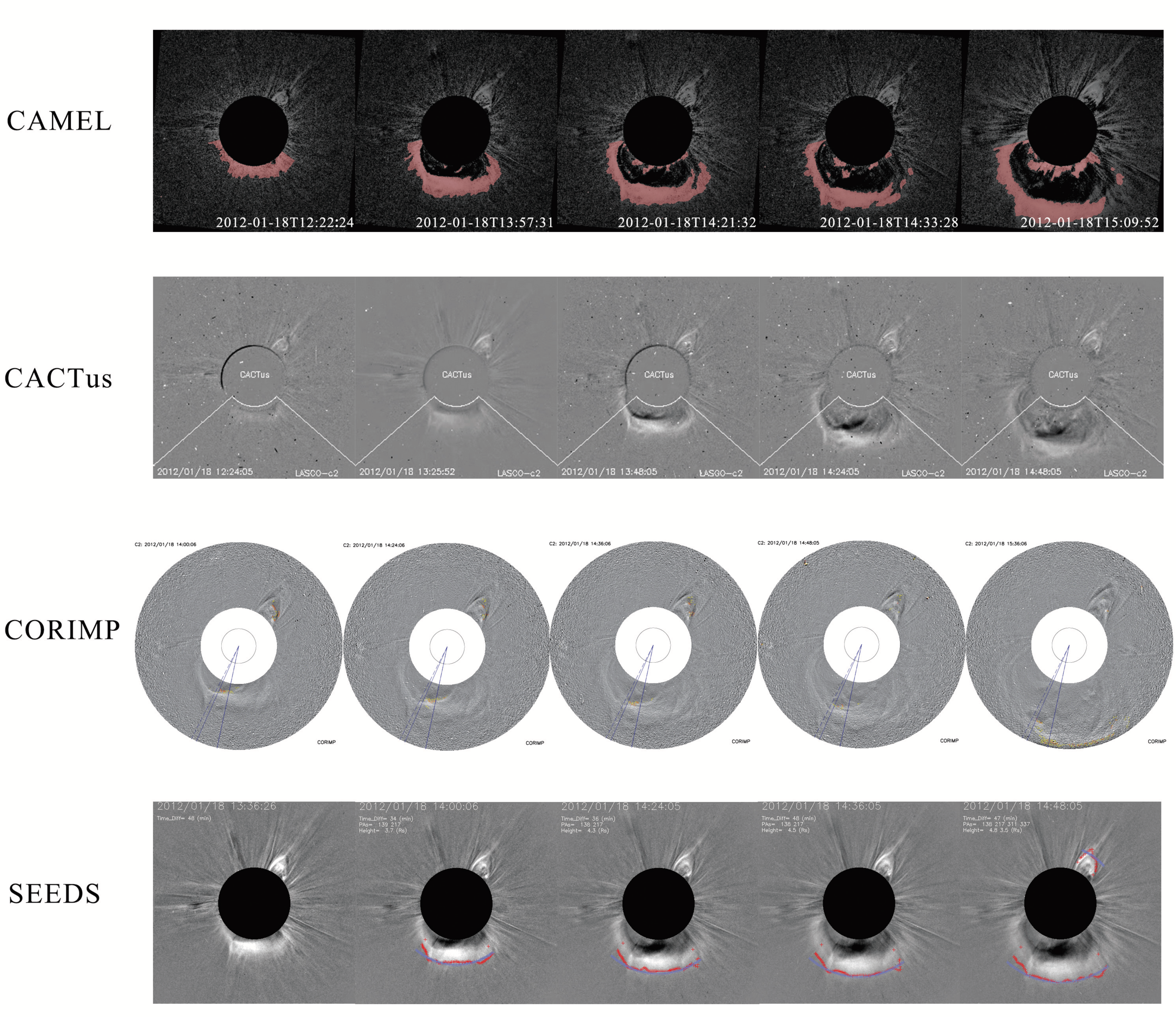}
	\caption{Comparison of the detection and tracking results for the CME event on 18 January 2012. }
	\label{fig:compare_3}
\end{figure}

\subsection{Partial halo CME on 18 January 2012}
When we get to halo CMEs, an important task is to identify a complete CME structure. Some frames illustrating the detection and tracking results with CAMEL, CACTus, CORIMP, and SEEDS are presented in Figure~\ref{fig:compare_3} from top to bottom. It reveals that CAMEL has a better performance to detect the partial CME structure as complete as possible, thanks to its capability of weak-signal detection. Both CACTus and SEEDS can only detect the northern bright segment of the CME, and fail to identify the weaker eastern part of the CME. CORIMP only outputs a very narrow region of the CME with the AW of 12 degrees. The AWs from the other four methods CAMEL, CACTus, SEEDS, and CDAW are 182, 98, 72, and 203 degrees, respectively. 
Consistently, all methods yield the CPA with a narrow distribution from 162 to 180 degrees. Concerning the first appearance time in LASCO C2, automatic CAMEL and CACTus identify the CME at the same time as the manual CDAW. CORIMP and SEEDS find the CME about 1.5 hours later. The CME under investigation is a rather slow CME with the velocity in the range of 241 to 363 km s$^{-1}$.

\begin{figure}[htbp]
	\centering
	\includegraphics[width=0.98\textwidth]{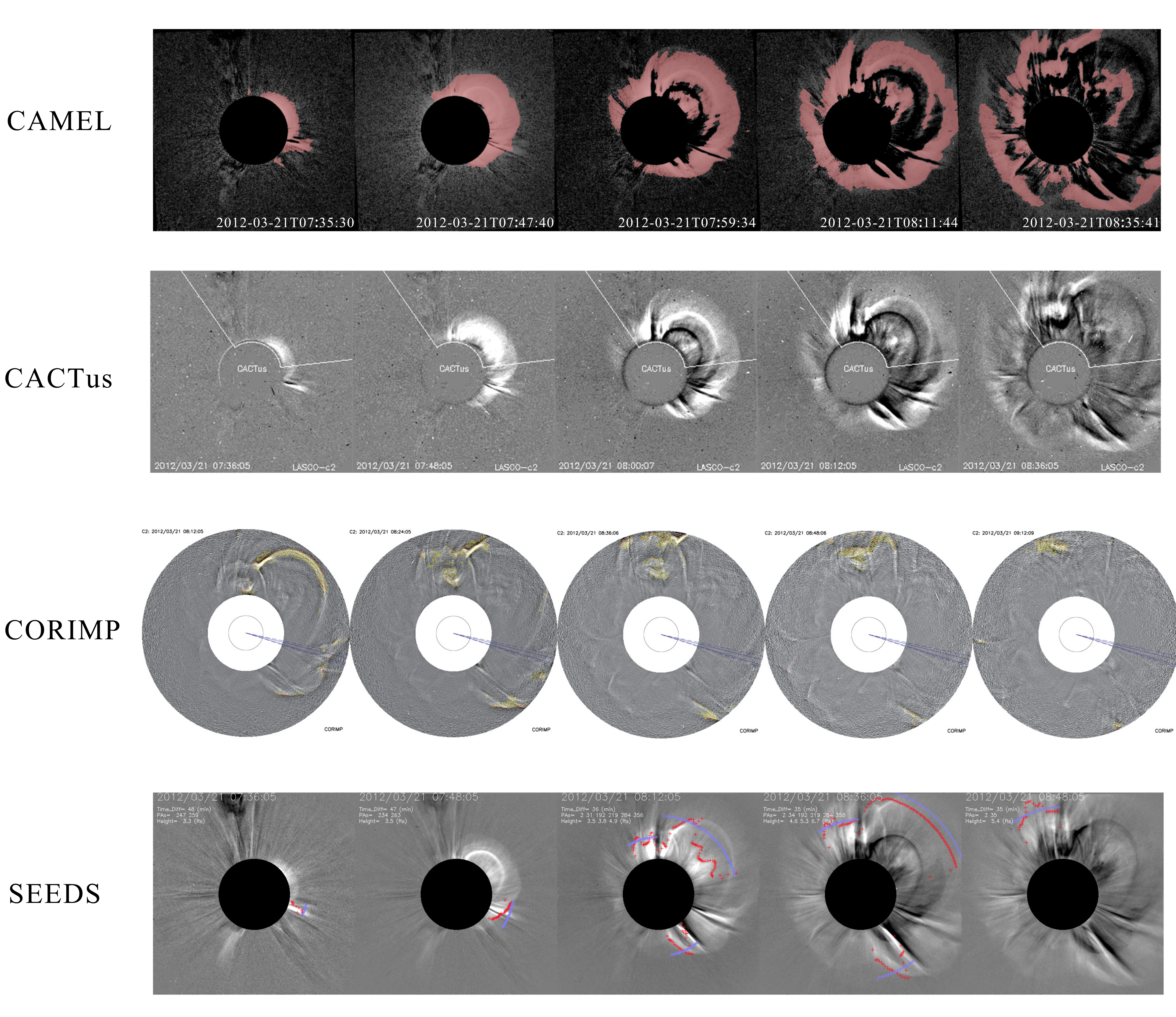}
	\caption{Comparison of the detection and tracking results for the CME event on 21 March 2012. }
	\label{fig:compare_4}
\end{figure}

\subsection{Full halo CME on 21 March 2012}
Full halo CMEs with 360-degree angular width belong to a category of CMEs which is mostly earth effective. If the CME is Earth-directed and carries southward magnetic fields, it usually causes a geomagnetic storm. Furthermore, if the CME is super-Alfv\^{e}nic, the driven shock wave may enhance the level of the geomagnetic storm. Therefore, the automatic detection and tracking of full halo CMEs is of particular importance for the purpose of space weather prediction. Note that the identified full halo CME might be actually a driven shock wave as we mentioned at the beginning of this Section.   

Figure~\ref{fig:compare_4} compares the detection and tracking results of the CME on 21 March 2012 obtained with CAMEL, CACTus, CORIMP, and SEEDS from top to bottom. Among the four techniques this event is firstly detected with our CAMEL method  at 07:35~UT which is the same as CDAW. CACTus, CORIMP, and SEEDS identify the CME at 07:48~UT, 08:00~UT, and 08:00~UT, respectively. Concerning the AW, only CAMEL is able to detect the whole CME (actually shock) structure. All the other three methods fail to identify especially the fainter shock wave segment on the left side of the deflected streamers. The corresponding AWs are 118, 3, 74 degrees for CACTus, CORIMP, and SEEDS, respectively.  In fact, SEEDS has two different entries for this CME in its catalog. In Table~\ref{table1}, only the entry with a larger AW is included.  The velocity increases from 812 km s$^{-1}$ to 1838 km s$^{-1}$ as obtained by SEEDS and CACTus. And CAMEL has a velocity of 1042 km s$^{-1}$ which is closest to the value of 1178 km s$^{-1}$ derived by CDAW. 

\section{Conclusions and Discussions}

We have implemented a novel system which can detect and track pixel-level CME regions in a set of white-light coronagraphs by using machine learning methods. The system is made of a three-module algorithm pipeline. The input of the system is a sequence of running-difference images observed by LASCO C2 and processed to level 1. In the first module, we use a well-trained supervised convolutional neural network to help classify whether a LASCO C2 image contains CME structures or not. The second module is to detect the pixel-level CME regions in the CME-flagged images by using an unsupervised PCA co-localization method, and to use graph cut method to refine the detection. The final module is to track each individual CME in the continuous CME image sequence with detected CME pixels produced by the first and second modules, respectively.  
The pipeline is run on a simple personal computer (PC) with a six-core i7 8700K processor, a NVIDIA gtx 1060 GPU graphic card with 6 GB frame buffer, and a computer memory of 8 GB. For the training of 10-month LASCO C2 data for image classification, it takes us about ten-hour computation on this PC. Training is a bit time consuming, but we need to only run it once after designing a proper training algorithm. The detection and tracking after the training is very fast. To process one-day LASCO C2 images of about 120 frames with $512\times512$ resolution, it takes about five to ten minutes to complete the task of detection and tracking.   

To evaluate the performance of our CAMEL technique, we select a few representative CME events which covers a large range of angular width, velocity and brightness, and compare our results with those obtained from a few existing CME catalogs. The available catalogs include manual CDAW catalog, automatic CACTus, SEEDS, and CORIMP catalogs. We have compiled a few fundamental parameters of CMEs for comparison, i.e., first appearance time in C2, central position angle, angular width, and velocities. Moreover, the detection and tracking results are also compared in a frame-by-frame manner. Couple of advantages we have seen for CAMEL is that:
\begin{itemize}
\item CAMEL is more qualified to detect and track weak signals to output more complete CME structures. Therefore, we can derive more precise CME morphological parameters, e.g., CPA, AW.
\item CAMEL has better performance to catch the appearance time of CMEs in LASCO C2 as early as possible. The more precise information of CME morphology and timing can eventually be used to derive more accurate CME kinematics.
\item CAMEL records pixel-level positions of CMEs. Thus we are detecting and tracking not only the CME leading front, but also any other interested structure in the detected CME regions, e.g., the CME core in Figure~\ref{fig:compare_2}.  
\item CAMEL has a by-product of the image classification. Categorizing CME-flagged images is very useful for distributing CME data timely and effectively to different space weather prediction center to predict CME arrivals. 
\item CAMEL is computationally cheap and fast. After a proper training, the detection and tracking of the CME in a single LASCO C2 image only takes a few seconds on a normal PC.  
\end{itemize}

This paper aims to introduce the detailed mathematical method of our newly developed tool for automatic detection and tracking of CMEs with the novel machine learning technique and evaluate its performance. Based on the automatic detection and tracking results, a chain of automatic processes can be carried out in future. We can automatically obtain the CME periphery from the pixel-level CME positions. Further 3D reconstructions of a CME based on the identified CME from single or multi-perspective observations can be done \citep[e.g.][]{Feng:etal:2012, Feng:etal:2013, Lu:etal:2017}. The 3D parameters derived from 3D reconstructions can be further used as inputs to MHD simulations, e.g., the ENLIL simulations of the CME propagation in the interplanetary space for the prediction of CME arrivals \citep{Odstrcil:2003}.
 
\acknowledgements
The CDAW CME catalog is generated and maintained at the CDAW Data Center by NASA and The Catholic University of America in cooperation with the Naval Research Laboratory. SOHO is a project of international cooperation between ESA and NASA.
This paper uses data from the CACTus CME catalog, generated and maintained by the SIDC at the Royal Observatory of Belgium.
We also acknowledge using the CME catalogs of SEEDS and CORIMP. This work is supported by NSFC grants (11522328, 11473070, 11427803, U1731241), by CAS Strategic Pioneer Program on Space Science, Grant No. XDA15010600, XDA15052200, XDA15320103, and XDA15320301, and National key research and development program 2018YFA0404202.

\bibliographystyle{aasjournal}
\bibliography{camel}
\end{document}